\documentstyle[12pt,a4wide]{article}

\def\e{{\rm e}}

\begin{document}

\bigskip

\bigskip

\bigskip

\centerline{\Large\bf ON BLACK HOLE STABILITY}
\vskip5mm
\centerline{\Large\bf IN MULTIDIMENSIONAL GRAVITY}
\vskip5mm
\centerline{ \bf U. Bleyer
\footnote{This work was supported by KAI e.V. grant WIP-016659},
    V.N. Melnikov\footnote{On leave of absense from Centre for Surface
     and Vacuum Research, Moscow. Work partly supported by the Russian
Ministry of Science and by DFG grant 436 RUS 113-7-2}    }
\centerline{\em WIP-Gravitationsprojekt, Universit\"at Potsdam }
\centerline{\em An der Sternwarte 16, D-14482 Potsdam, Germany }
\centerline{e-mail: ubleyer@aip.de }
\vskip5mm
\centerline{\bf K.A.Bronnikov and S.B.Fadeev}
\vskip5mm
\centerline{\em Centre for Surface and Vacuum Research,}
\centerline{\em 8 Kravchenko str., Moscow, 117331, Russia}
\centerline{e-mail: mel@cvsi.uucp.free.msk.su}
\vskip6mm
\noindent
     ABSTRACT
\vskip3mm
\noindent
     Exact static, spherically symmetric solutions to the
     Einstein-Maxwell-scalar equations, with a dilatonic-type scalar-vector
     coupling, in $D$-dimensional gravity with a chain of $n$ Ricci-flat
     internal spaces are considered. Their properties and special cases are
     discussed. A family of multidimensional dilatonic black-hole
     solutions is singled out, depending on two integration constants
     (related to black hole mass and charge) and three free parameters of
     the theory (the coordinate sphere, internal space dimensions, and the
     coupling constant). The behaviour of the solutions under small
     perturbations preserving spherical symmetry, is studied. It is shown
     that the black-hole solutions without a dilaton field are stable,
     while other solutions, possessing naked singularities, are
     catastrophically unstable.

\vfill
\pagebreak
\section{INTRODUCTION}
With this paper we continue a series of investigations devoted to
multidimensional gravitational models with the aim to obtain exact
solutions and to find on their basis some observational consequences
of extra (hidden) dimensions in our physical world. We also study the
role of different physical fields in a self-consistent manner.

     Modern theories of field unification assume in general a greater
space-time dimensionality than four. Although multidimensional gravity
as an approach to field unification can  be traced back  to  the
famous works of  Kaluza  and Klein [1,2] in the twenties, today's
increased interest to this field is largely stimulated by studies in
supersymmetry and some other modern theories [3]; in the
field-theoretical limit of such theories gravity is described with
reasonable accuracy by multidimensional Einstein equations.  Studies
of their solutions can lead to predictions of direct observational
manifestations of extra dimensions. Thus, cosmological models predict
variations of the gravitational constant $G$, so that observational
constraints imply certain limits on model parameters [4,5]. In these
papers exact solutions as well as relations between the rate of a possible
change of $G$ with the Hubble parameter, the deceleration parameter
and the mean density of the Universe were obtained. Another
possible window to the multidimensional world is opened by an analysis of
local effects which could be sensitive to spatial variations of
extra-dimension parameters [6]. We shall discuss some effects of
this sort, in particular, those connected with electric charges of
isolated bodies [6-8]. Quantum variants of all these models [4] deal
with problems of quantum wormholes, creation of the Universe via
tunnelling, nonsingular models etc.

Here, we consider exact, static, spherically symmetric solutions to
the Einstein-Maxwell-scalar field equations, with a dilatonic type
coupling between the scalar and vector fields, in $D$-dimensional
gravity with a chain of $n$ Ricci-flat internal spaces and an
arbitrary dimension of coordinate spheres (orbits of the group of
spatial motions) --- see Eq.(6) for the space-time structure.  In
particular, we pay attention to the arbitrariness of $d$, in the
spirit of Tangherlini's generalization of the Schwarzschild metric
[9], with the hope to shed some light at the role of the 4-dimensional
nature of the physical space-time.

     Our approach differs from that applied in some papers on
multidimensional black holes in that any solutions, not only
black-hole ones, are sought. Consequently, the place of black holes
(if any) in the whole set of solutions, the role of fields for the
existence of different types of solutions, as well as the properties
of all spherical configurations, become clearer. Here, we call a black
hole solution a configuration with the central singularity screened by
an event horizon.  In addition, we study the stability of these
solutions under small perturbations preserving spherical symmetry and
arrive at a result of physical interest that from the whole set of
static solutions only the black-hole ones are stable.


\section{BASIC EQUATIONS AND THE GENERAL \newline  STATIC SOLUTION}

     Let us consider a system with the Lagrangian
\begin{equation}                                                       
     L=R^{(D)}+g^{MN}\varphi_{'M}\varphi_{'N}
               -\e^{2\lambda \varphi}F^{MN}F_{MN}
\end{equation}
     for the set of interacting fields arising in the field limit of
superstring theories. Here $\varphi$ is a dilaton scalar field and
$F_{MN}=\partial_M A_N - \partial_N A_M$ is an Abelian gauge field
interpretable as the electromagnetic one. As pointed out in Ref.[10],
an electromagnetic field introduced in the multidimensional action may
seem less aesthetic than a purely gravitational (Einstein) action
usually considered in Kaluza-Klein theories but it appears that
elementary gauge fields are necessary also in higher dimensions in
order to obtain a realistic grand unification theory.

     The corresponding set of field equations is
\begin{eqnarray}                                                       
 \nabla^M \nabla_M \varphi +\lambda \e^{2\lambda\varphi}F^{MN}F_{MN}&=&0,\\
         \ \ \nabla_N(\e^{2\lambda\varphi}F^{NM})&=&0, \\                 
     R_{MN}-g_{MN}R^A_A/2 &=& -T_{MN}                                    
\end{eqnarray}
     where $T_{MN}$ is the energy-momentum tensor
\begin{equation}                                                        
     T_{MN} = \varphi_M\varphi_N - \frac{1}{2}g_{MN}\varphi^A\varphi_A +
    \e^{2\lambda\varphi}[-2F_M^A F_{NA}+\frac{1}{2}g_{MN}F^{AB}F_{AB}].
\end{equation}
     Capital Latin indices range from 0 to $D-1$, the gravitational
constant is put equal to unity.

     Let us choose a $D$-dimensional manifold of the structure
\begin{equation}                                                       
   M=M^{(3+d)}\times M_1\times \cdots \times M_n; \ \ \dim M_i=N_i; \ \
          D=3+d+\sum_{i=1}^{n}N_i,
\end{equation}
     with the signature $(+, -, \ldots, -)$,
     where $M^{(3+d)}$ plays the role of the ordinary space-time and $M_i$
     are Ricci-flat manifolds with the line elements
     $ds_{(i)}^2$, $i=1,\cdots, n$.  We seek solutions of the field equations
     such that $M^{(3+d)}$ is a static, generalized
     spherically symmetric space-time with the metric
\begin{equation}                                                      
     ds_{d+3}^2 =
     \e^{2\gamma(u)}dt^2-\e^{2\alpha(u)}du^2 - \e^{2\beta(u)}d\Omega_{d+1}^2
\end{equation}
     where $d\Omega_{d+1}^2$ is the line element on a unit
     $(d+1)$-dimensional sphere $S^{d+1}$,
     while all the scale factors $\e^{\beta_i}$ of the internal spaces
     $M_i$ depend on the radial coordinates $u$, i.e., the $D$-metric is
\begin{equation}                                                      
      ds_D^2 = g_{MN}dx^M dx^N=
          ds_{d+3}^2 - \sum_{i=1}^{n} \e^{2\beta_i(u)}ds_{(i)}^2.
\end{equation}
     If we denote $\gamma =\beta_{-1}, N_{-1}=1,  \beta  =\beta_0,  N_0=2$ and
     choose the harmonic radial coordinate $u$ such that
\begin{equation}                                                      
     \alpha = \sum_{i=-1}^{n}\beta_iN_i.
\end{equation}
     the Ricci tensor components $R_M^N$ can be written in the highly
     symmetric form ($x^1=u$)
\begin{eqnarray}                                                      
     R_1^1 &=& -\e^{-2\alpha}\sum_{i=-1}^{n}N_i [\beta_i''
       +\beta_i^{'2}-\beta_i'\alpha']; \nonumber \\
     R_{\mu}^N &=& 0 \ \ \ (N>d+2;\ \ \mu =0,\cdots ,d+2); \nonumber \\
     R_{a_j}^{b_i} &=&
     -\delta_j^i\delta_{a_i}^{b_i}\e^{-2\alpha}\beta_i'', \ \
                    i\ne 0      \nonumber\\
     R_2^2 &=& \ldots= R_{d+2}^{d+2} = (d+1)\e^{-2\beta}
          -\e^{-2\alpha}\beta'';
\end{eqnarray}
     where the indices
     $a_j(b_i)$ refer to the subspace $M_j(M_i)$.

     Let us further assume the dilaton scalar field to be
     $\varphi=\varphi(u)$ and the electromagnetic field
     to be Coulomb-like:
     $A_M=\delta_M^0 A_0(u)$. Then the $D$-dimensional Maxwell-like equations
     give:
\begin{eqnarray}
     F^{01} = q\e^{-2\lambda\varphi}/\sqrt{g} =
  q\e^{-2\lambda\varphi-2\alpha}, \ \ q={\rm const\ (charge)},\nonumber \\
      g =|\det\,g_{MN}| =\exp\bigl(2\alpha
           +2\sum_{i=-1}^{n}N_i\beta_i\bigr) = \e^{4\alpha}.
\end{eqnarray}                                                         
     (the metric determinants of $ds_{(i)}, i=1,\ldots,n$, never appear in
     the equations and may be omitted).

     Now the scalar and (some linear combinations of) the metric field
     equations may be written in the form
\begin{eqnarray}
     \varphi''+2q^2\lambda \e^{2\gamma-2\lambda\varphi}&=&0;\\       
     N\beta_i''+\gamma''&=&0, \ \ \ \ i=1,\ldots,n; \ \ N=D-3;\\      
     (\alpha-\beta)''-d^2 \e^{2\alpha-2\beta} &=&0;           \\     
     \gamma''-\frac{2q^2 N}{N+1}\e^{2\gamma-2\lambda\varphi} &=&0    
\end{eqnarray}
\begin{equation}
     \alpha^{'2} - \sum_{i=-1}^{n}\beta_i^{'2} -d(d+1)\e^{2\alpha-2\beta}=
               \varphi'^2 - 2q^2 \e^{2\gamma-2\lambda\varphi}          
\end{equation}
     where Eq.(16) is the $(^u_u)$ component of the Einstein equations and
     forms a first integral of the remaining equations (12-15).

     Eqs.(13, 14)  are easily integrated to give
\begin{eqnarray}
   \beta_i = -\gamma/N + h_i u,\ h_i ={\rm const,}\ i=1,\ldots,n; \\  
    \e^{\beta-\alpha} = d\cdot s(k,u), \    k ={\rm const}            
\end{eqnarray}
    where for any choice of variables
\begin{equation}                                                      
     s(a,x) = \left\{\begin{array}{ll}
           a^{-1}\sinh\,ax, \ & a>0; \\
           x,                 & a=0; \ \ \ \ \\
           a^{-1}\sin\,ax,    & a<0.  \end{array} \right.
\end{equation}
     and inessential integration constants
     have been eliminated by shifting the origin of $u$ and rescaling the
     coordinates in the subspaces $M_i$ (their true scales are thus hidden
     in $ds_{(i)}^2$).

     Certain combinations of Eqs.(12) and (15) are also easily integrated
     giving
\begin{eqnarray}
     \varphi &=& Cu/A -2\lambda N_{+}\omega,\\                      
     \gamma &=& (\omega +\lambda Cu)/A,                            
\end{eqnarray}
     with the function $\omega(u)$ defined by
\begin{equation}
     \e^{-\omega}=Qs(h,u+u_1); \ h,u_1={\rm const}; \ Qs(h,u_1)=1   
\end{equation}
     where  $h$ and $C$ are integration constants; the other
     constants are defined by
\begin{equation}                                                      
     N=D-3;\ \  A = 1+\lambda^2(N+1)/N, \ \ Q^2=q^2/N_{+},\ \
          N_+=(N+1)/(2AN).
\end{equation}
     Since, by our choice of the origin of $u,\ \e^{\beta}\rightarrow\infty$
     for $u \rightarrow 0$, the value $u=0$ corresponds to spatial infinity.
     The last condition from (22) is the requirement that $\gamma =0$ at
     infinity, i.e., $dt$ is a time interval measured by a distant observer's
     clock.

     The final form of the $D$-dimensional metric is
\begin{equation}                                                      
     ds_D^2=\e^{2\gamma}dt^2 -
 \e^{-2\gamma/N}\left\{\left[\frac{\e^{-Bu}}{d\cdot s(k,u)}\right]^{2/d}
     \left[\frac{du^2}{d^2 s^2 (k,u)}+d\Omega_{d+1}^2\right]
     +\sum \e^{2h_i u}ds_{(i)}^2\right\},
\end{equation}
\begin{equation}                                                      
     \e^\gamma=\left[\frac{\e^{\lambda Cu}}{Qs(h,\,u+u_1}\right]^{1/A}.
\end{equation}
     Here and henceforth $\sum = \sum_{i=1}^{n}$.
     The integration constants are connected by the relation
\begin{equation}                                                      
     \frac{d+1}{d}k^2{\rm sign}k = 2N_{+}h^2{\rm sign}h +
     \frac{C^2}{A}+\frac{B^2}{d}+\sum N_i h_i^2.
\end{equation}
     This general static spherically-symmetric solution has $(n+3)$
     essential integration constants: the scalar charge $C$, the electric
     charge $q$ (or $Q$), the ``charges'' of extra dimensions $h_i$ and
     one of the constants $h$ or $k$ connected by Eq.(26). For $d=1$ (i.e.,
     conventional spherical symmetry with 2-dimensional spheres) $h$ along
     with $C$ and $Q$ can be related to the mass $m$ which is defined by
     comparing the asymptotic $u\rightarrow 0 \ \ (r\rightarrow \infty )$
     of $\e^{\gamma}$ with the Schwarzschild metric:
\begin{equation}                                                      
     AGm+\lambda C=(Q^2+h^2 {\rm sign} h)^{1/2},
\end{equation}
     The coordinate $u$ is defined in the domain $[0,\infty )$ if $h\geq
     0,u_1>0$ or between zero and some $u_{\max}>0$ in other cases. The
     scale factors $\e^{\beta_i}=1$ for $u=0$.


\section{PROPERTIES OF THE SOLUTION \newline AND SPECIAL CASES}

     Let us consider some special cases of the solution.
\begin{description}
\item[(a)]
     The electromagnetic field is eliminated when $Q=0$. Then we recover the
     generalized [7] Tangherlini solution [9]; the transformation
\begin{equation}                                                 
     \e^{-2ku}= 1-\frac{2k}{R} \equiv f(R);
      \ \ \frac{-h+\lambda C}{A}=ka;
      \ \ h_i=k\left(-a_i+\frac{a}{N}\right),
\end{equation}
     brings it to the form                                          
\begin{eqnarray}
     ds_D^2 &=& f^a dt^2 -[R^2f^{1-a-b}]^{1/d}\left[\frac{dR^2}{R^2 f}+
               d\Omega_{d+1}^2\right]
               - \sum_{i=1}^{n}f^{a_i}ds_{(i)}^2, \nonumber \\
       \varphi &=&-\frac{\overline{C}}{2k}\ln f(R), \ \
               \overline{C}=\frac{1}{A}(C+2\lambda N_{+}h), \nonumber \\
       (d+1)/d &=& \sum N_i a_i^2+a^2+ (a+b)^2/d + \overline{C}^2/k^2
\end{eqnarray}
     where $b = \sum N_i a_i$. The
     $(d+3)$ dimensional part of (29) coincides with the Schwarzschild
     solution when
     $d=1$ (the physical space-time is 4-dimensional) and $h_i =0$.
\end{description}

     The scalar field in (29) affects the metric only via the constant
     $\overline{C}$ in the relation among the constants.

     The space-time (29) has a horizon (at $r=2k$) only in the
simplest case $ a-1 = a_i =b = \overline{C} = 0$. Thus in our model an
electrically neutral black hole can exist only with ``frozen'' extra
dimensions ($\e^{\beta_i}={\rm const}$) and with no scalar field.
(The latter result is well known in conventional general relativity: a
massless, minimally coupled scalar field is incompatible with an event
horizon [11]). In this sense the generalized Schwarzschild-Tangherlini
black holes may be called trivial. These are examples of the so called
''spontaneous''compactification of extra dimensions.

\begin{description}
\item[(b)]
     when $\lambda =0$, we obtain the generalized
     Reissner-Nordstr\"om-Tangherlini solution for linear scalar and
     electromagnetic fields discussed in Ref.[8]. \\
\item[(c)]
     The scalar field is ``switched off'' when $\lambda =C=0$, which
     yields a special case of item (b).
\end{description}
     A new feature implied by a nonzero electric charge as compared with item
     (a) is that the constants $k$ and $h$ can have any sign and the
     function $s(h, u+u_1)$ can be sinusoidal, which yields
     $u_{\max}<\infty$. Physically that means the appearance of a
     Reissner-Nordstr\"om repelling singularity at the center of the
     configuration.
\begin{description}
\item[(d)]
     When extra dimensions are absent, we arrive at a solution for
     interacting scalar and electromagnetic fields in (d+3)-dimensional
     general relativity; for the conventional case $d=1$ it was obtained in
     Ref.[12] (s. also [14]).\\
\item[(e)]
     For the case of 4-dimensional physical space-time (d=1, i.e., in
     Eq.(5) the space $M^{3+d}=M^4 = R^1 \times R^1 \times S^2$),
     the corresponding solution was obtained in Refs.[6,7,13].
\end{description}

     Comparing the solutions with and without extra dimensions, one sees that
     indeed the scale factors $\beta_i$ of the Ricci-flat spaces $M_i$ behave
     like additional minimally coupled scalar fields.

     For nonzero electric charge, i.e., in the general case, there is
no choice of integration constants such that the extra dimensions are
frozen ($\beta_i=const$). So, in this case we have no solutions with
spontaneous compactification. The behaviour of the metric coefficients
for different combinations of integration constants is rather various:
thus, for $u\rightarrow \infty$ (provided $h \geq 0$)
\begin{equation}                                        
\beta\sim \left[\frac{-h+\lambda C}{AN}-B-k\right]u, \ \ \gamma
\sim \frac{-h+\lambda C}{A}u, \ \ \beta_i\sim\left[\frac{-h+\lambda
C}{AN}+h_i\right]u,
\end{equation}
so they may be finite or infinite.
Calculations show that the solution has a naked singularity at
$u=u_{\max}$ or $u=\infty$ in all cases except
\begin{equation}                                         
h_i=-k/N; \ \ \ h=k; \ \ \ C=-\lambda k(N+1)/N,
\end{equation}
when the sphere $u=\infty$ is a Schwarzschild-like event horizon: at
finite radius $r=\e^{\beta}$ of a coordinate sphere the integral $\int
\exp(\alpha -\gamma )du$ for the light travel time diverges. So, (31)
is the black hole solution we are looking for.

     In general, four types of field behaviors may be singled out:
\begin{enumerate}
\item[A.] If $h<0$ or (and) $u_1<0$, which may take place only when there is a
     nonzero charge $q$, the coordinate $u$ is defined between 0 and a
     certain finite value $u_{\max} = {\rm zero}$. \\$ s(h, u+u_1) < \infty$
     corresponds to a repelling (since $g_{00}\rightarrow\infty$) naked
     Reissner-Nordstr\"om-like singularity where the electric field tends to
     infinity while the scalar field $\varphi$ and the extra-dimension scale
     factors $\beta_i$ remain finite.
\item[B.] The case $h>0, \ u_1 >0;\; r\rightarrow 0$ as $u\rightarrow\infty$.
     The value $u\rightarrow\infty$ corresponds to an
     attracting ($g_{00}\rightarrow 0$ scalar-field dominated naked central
     singularity where the dilaton field $\varphi$ and the ``scalar fields''
     $\beta_i$ (at least some of them) become infinite.
\item[C.] The case $h>0, \ u_1 >0;\; r\rightarrow \infty$ as
     $u\rightarrow\infty$. This configuration, sometimes called a space
     pocket (``Raumtasche'' by P.Jordan), possesses an attracting naked
     singularity located at a sphere of infinite radius hidden beyond a
     regular ``throat'', i.e., a minimum of the coordinate sphere radius
     $\e^{\beta}$ which is reached at some finite value of the $u$
     coordinate. Again at least some of the ``scalar fields'' are infinite at
     the singularity. The system geometry would be like that of a wormhole if
     it were regular at $u=\infty$.
\item[D.] The special case (31), which is intermediate between cases B and C,
     corresponds to a configuration called a dilatonic black hole. In
     General Relativity ($D=4, d=1$) such a solution was
     obtained for the first time in Ref.[12] (see also [14]) and was recently
     widely discussed in various aspects (see, e.g. [15] and references
     therein).  We will also pay some attention to it.
\end{enumerate}

     In the black-hole case (31) only two independent integration constants
     remain, say,  $k$ and $Q$, and the transformation (28) brings the
     solution to the form
\begin{eqnarray}                                                    
     ds_D^2&=
     &\frac{(1-2k/R)dt^2}{(1+p/R)^{2/A}}                \nonumber\\
     &&-(1+p/R)^{2/AN}\left[\left(\frac{R^{1-d}}{d^{d+1}}\right)^{2/d}
     \frac{dR^2}{1-2k/R}+\left(\frac{R}{d}\right)^{2/d} d\Omega_{d+1}^2
          +\sum_{i}ds_i^2\right], \\
     \varphi &=& \frac{\lambda}{A}\frac{N+1}{N}                     
          \ln (1+p/R),\\
     p&=&\sqrt{Q^2+k^2}-k.                                          
\end{eqnarray}
     It is a generalization of the Reissner-Nordstr\"om solution.
     For $d=1$ the metric takes the form
\begin{equation}                                                      
     ds_D^2
     =\frac{(1-2k/R)dt^2}{(1+p/R)^{2/A}}
     -(1+p/R)^{2/AN}\left[\frac{dR^2}{1-2k/R}
     +R^2d\Omega^2+\sum_{i}ds_i^2\right]
\end{equation}
     considered in Refs. [16] (in other notation) and [13] and is in turn
     reduced to the genuine Reissner-Nordstr\"om metric if one puts, in
     addition, $D=4$ (no extra dimension), $\lambda =0$ (no scalar
     field).  For $d=1$, analysing the asymptotic $R\rightarrow\infty$, one
     naturally introduces the so-called Schwarzschild mass $m$ expressible in
     terms of $Q$ and $k$ (or $k$ is expressed in terms of $Q$ and $m$):  now
     Eq.(34) is supplemented by $p=A(Gm-k)$.

     In this family of black-hole solutions a nonzero scalar dilaton
     field (33) exists solely due to the interaction ($\lambda\ne 0$).
     When $\lambda=0$, i.e., the $\varphi$ field becomes minimally coupled, a
     horizon is compatible only with $\varphi=$const. This conforms with the
     well-known ``no-hair'' theorems and the properties of the
     general-relativistic scalar-vacuum and scalar-electrovacuum
     configurations [11,14].

     It should be pointed out that the solution (32-34) is a very special
     case of the general solution (2 integration constants instead of
     $(n+3)$).  So there should exist very strong arguments for an assertion
     that in multidimensional gravity an actual collapse of a spherical
     body should lead to black hole (BH) formation.  Such arguments do arise
     when we investigate the stability of the static solutions.

     It can be demonstrated that one of the specific potentially observable
     effects for multidimensional systems is the  violation
     of the Coulomb law.
     Indeed, the electric field strength $E=(F^{01}F_{10})^{1/2}$
     reads for
     our general solution (10) in the physical case $d=1$
\begin{equation}                                                      
     E=(|q|/r^2)\e^{-\sigma-2\lambda \varphi }
\end{equation}
     where $r(u)=\e^{\beta}$ is (as before) the curvature radius of a
     coordinate sphere (the curvature coordinate) and
\begin{equation}                                                      
     \sigma \equiv \sum_{i=1}^{n}N_i\beta_i.
\end{equation}
     As seen from Eq.(35), the deviations from the Coulomb law are both due
     to extra dimensions and due to the scalar-electromagnetic interaction.

     In the BH case for $d=1$ from the asymptotic $(r\rightarrow \infty )$
     of the metric (35) we have:
\begin{equation}                                                      
     E=\frac{|q|}{r^2}\left\{1-\frac{1}{r}\left[(Gm-k)\frac{N-1}{N} +
2\lambda^2(Gm+k)\frac{N+1}{N}\right]+O\left(\frac{1}{r^2}\right)\right\}
\end{equation}
     One can conclude that the magnitude of Coulomb law violation  is of the
     order of the gravitational field strength characterized by the ratio
     $Gm/r$ and depends also on $N$ and $\lambda$.

\section{STABILITY ANALYSIS}

     Now we would like to investigate the stability of our static solutions
     under perturbations preserving spherical symmetry (i.e., monopole
     modes).

     As verified by experience, although the D-dimensional setting of the
     problem is quite convenient for finding the static solutions, it makes a
     stability investigation very awkward. The latter is better carried out
     in a $(d+3)$-dimensional formulation.

     {}From the viewpoint of the physical space-time $V_{d+3}$ the
     scale factors $\beta_i$ are additional scalar fields. Thus in the
     Lagrangian (1) the $D$-curvature and the metric are to be written
     explicitly in terms of $\beta_i$, after which the Lagrangian acquires
     the characteristic form for a scalar-tensor theory of gravity:

\begin{equation}                                                        
     L=\e^{\sigma}\left[R^{(d+3)}-\sigma^{\mu}\sigma_{\mu}
          +\sum \beta_{i,\mu}\beta_i^{\mu} + \varphi^{\mu}\varphi_{\mu}
          -\e^{2\lambda\varphi} F^{\mu\nu}F_{\mu\nu}\right]
\end{equation}
     where $R^{(d+3)}$ corresponds to the metric $g_{\mu\nu},\ \sigma$ is
     defined in (37) and it is assumed that
$$
     \varphi = \varphi(x^{\mu}), \ F_{\mu\nu} = F_{\mu\nu}(x^{\alpha}); \
     F_{MN}=0 \ {\rm for} \ M \ {\rm or}\  N>d+2.
$$
     The factor $\e^{\sigma}$ emerges from the $D$-dimensional determinant.

     It is helpful to pass from $g_{\mu\nu}$ to the conformal metric
     $\overline{g}_{\mu\nu}$ defined by
\begin{equation}                                                        
     \overline{g}_{\mu \nu} = \e^{-2\sigma/(d+1)}g_{\mu \nu}
\end{equation}
     The Lagrangian acquires the form
\begin{equation}                                                        
     \overline{L}=\overline{R}^{(4)}+
     \frac{1}{d'}\sigma^{,\alpha}\sigma_{,\alpha}
     +\sum N_i \beta_{i,\mu}\beta_i{}^{,\mu}+
     \varphi^{,\mu}\varphi_{,\mu}
     -\e^{2\sigma/d' +2\lambda\varphi}F^{\alpha \beta}F_{\alpha \beta},
\end{equation}
     where  we have denoted $d'=d+1$, $\overline{R}^{(d+3)}$ corresponds to
     $\overline{g}_{\mu \nu}$ and the indices are raised and lowered using
     $\overline{g}_{\mu \nu}$.

     Taking $\overline{g}_{\mu \nu}$ in the standard static,
     spherically symmetric form
\begin{equation}                                                        
     ds_{d+3}^2 = \overline{g}_{\mu \nu}dx^{\mu}dx^{\nu} =
     \e^{2\overline{\gamma}}dt^2-\e^{2\overline{\alpha}}du^2
     -\e^{2\overline{\beta}}d\Omega_{d+1}^2
\end{equation}
     and choosing the harmonic coordinate $u$, i.e., putting
     $\overline{\alpha}= (d+1)\overline{\beta}+\overline{\gamma}$, one
     rather easily restores the general static solution (20)-(24). It is
     noteworthy that the coordinate $u$ is harmonic with respect to both the
     $D$-dimensional metric $g_{MN}$ and the $(d+3)$-dimensional metric
     $\overline{g}_{\mu\nu}$, i.e., $\Box_D u = \bar{\Box}_{d+3} u
     =0$, but not with respect to the $(d+3)$-dimensional part $g_{\mu\nu}$
     of the $D$-metric.

     The explicit form of the metric (42) is easily obtained from (24), (25)
     using the transformation (40).

     Now let us investigate small deviations from the
     static configurations
\begin{equation}                                                      
     \delta \varphi (u,t), \ \ \delta \beta_i(u,t), \ \
     \delta \overline{g}_{\mu \nu}(u,t), \ \ \delta F_{\mu \nu}(u,t),
\end{equation}
     preserving spherical symmetry, i.e. monopole modes. Thus
     dynamical degrees of freedom are restricted to the scalar field and
     the scale factors $\beta_i$ which in the 4-dimensional representation
     behave as effective scalar fields. For simplicity let us assume that
     there is only one internal space:
\begin{equation}                                                      
      N_1=D-d-3=N-d>0; \      \ N_2=N_3=\cdots =0.
\end{equation}

     In what follows, with no risk of confusion, we will omit the overbars at
     the symbols $\alpha, \beta, \gamma$.

     The next step is to choose the frame of reference and the
     coordinates in the
     perturbed space-time. Evidently this choice may be carried out by
     prescribing certain relations among the perturbations. Following [17],
     we would like to choose the so-called central frame of reference,
     where coordinate spheres of fixed radii are at rest, and the radial
     coordinate is taken such that the numerical values of these radii
     are the same as
     those in the static background configuration with the metric (42). Thus
     we postulate $\delta\beta \equiv 0$, or, as it is sometimes more
     convenient to use the radius $r\equiv\e^\beta$ (coinciding with the
     Schwarzschild radial coordinate), $\delta r \equiv 0$.

     The perturbed metric functions $\tilde{\gamma}(u,t)$ and
     $\tilde{\alpha}(u,t)$ are taken in the form
\begin{equation}                                                       
     \tilde{\gamma}(u,t)=\gamma (u)+\delta \gamma (u,t); \ \
     \tilde{\alpha} (u,t)=\alpha (u)+\delta \alpha (u,t),
\end{equation}
     and similarly for $\tilde{\varphi}(u,t),
     \tilde{\beta}_1(u,t)\equiv \tilde{\mu}(u,t)$ and
     $\tilde{F}_{\mu \nu}(u,t)$.  The perturbed Maxwell-like field is defined
     by $\tilde{A}_0(u,t)$.

     Integrating Eq.(3), we get
\begin{equation}                                                       
     \tilde{F}^{\alpha \beta}\tilde{F}_{\alpha\beta}
     =-2q^2\\e^{-4\lambda\tilde{\varphi}-2(N-1)\tilde{\mu}}r^{-2d'}
\end{equation}
     where $q$ is unperturbed since we study only dynamical perturbations
     rather than changes of integration constants and $r$ is unperturbed by
     the choice of the frame of reference.  From the metric field equations
     we obtain equations for $\delta \mu$ and $\delta \varphi$:
\begin{eqnarray}                                                       
     &r^{2d'}\delta \ddot{\mu}-\delta \mu''-\mu'(\delta \gamma'
          -\delta\alpha') + 2\mu''(\delta \alpha - w) & =0, \\
     &r^{2d'}\delta \ddot{\varphi}-\delta \varphi'' -\varphi'(\delta  
   \gamma'-\delta \alpha')+ 2\varphi''(\delta \alpha - w) & =0,
\end{eqnarray}  \begin{equation}
    w \equiv \lambda \delta \varphi + (N-d)\delta \mu/d'           
\end{equation}
     where $\mu',\mu'',\varphi'$ and $\varphi''$ are static functions
     {}from the background solution and
     $\delta \gamma'$ and $\delta \alpha$ are defined in (45).
     The $\left({1}\atop{0}\right)$-component of the metric field equation is
     easily integrated in $t$:
\begin{equation}                                                      
     d'\beta'\delta \alpha = \frac{(N-d)(N+1)}{d'}\mu'\delta \mu
     +\varphi'\delta \varphi + F(u),
\end{equation}
     and the difference of the $\left({0}\atop{0}\right)$ and
     $\left({1}\atop{1}\right)$ components gives:
\begin{equation}                                                      
     \frac{d'}{2}\beta'(\delta\alpha'+\delta\gamma')
     = \frac{(N-d)(N+1)}{d'}\mu'\delta\mu + \varphi '\delta \varphi .
\end{equation}
     Taking $\delta \alpha$ and $\delta \gamma'$ from (50) and (51) and
     substituting them into (47) and (48), we arrive at coupled
     wave equations for $\delta \mu$ and $\delta \varphi$:
\begin{eqnarray}                                                         
     r^{2d'}\delta \ddot{\varphi}-\delta \varphi''
     +\frac{2}{d'}\left[\left(\frac{r\varphi'^2}{r'}\right)'\delta\varphi
     +\frac{(N-d)(N+1)}{d'}\left(\frac{r\varphi'\mu'}{r'}\right)'\delta\mu
     \right] &=& 2\varphi''w;          \\
     r^{2d'}\delta \ddot{\mu}-\delta \mu''
     +\frac{2}{d'}\left[\left(\frac{r\mu'\varphi'}{r'}\right)'\delta\varphi
 +\frac{(N-d)(N+1)}{d'} \left(\frac{r\mu'^2}{r'}\right)'\delta\mu
     \right] &=& 2\mu''w.
\end{eqnarray}                                                           

     Our static system is unstable if there exist
     physically allowed solutions to Eqs.(52) and (53)
     growing at $t\rightarrow \infty$. A separate problem is to define which
     solutions should be accepted as physically allowed ones. Let us join the
     approach of Ref.[17] and require
\begin{equation}                                                        
     \delta \mu \rightarrow 0, \ \ \
     \delta \varphi \rightarrow 0 \ \ \ {\rm for} \ \ \  u\rightarrow 0
\end{equation}
     at spatial infinity $r\rightarrow \infty$ and
\begin{equation}                                                         
     \mid \delta \mu /\mu \mid < \infty,
     \ \ \mid \delta \varphi /\varphi \mid <\infty
\end{equation}
     at singularities and horizons. These requirements provide the validity
     of linear perturbation theory over the whole space-time, including the
     neibourhood of the singularities. We must in addition forbid energy
     fluxes to our system from outside; however [17], such a requirement
     constrains only the signs of the integration constants and does not
     affect any conclusions.

     There are some cases when the set of equations (52), (53)
     simplifies, i.e., decouples or reduces to a single
     equation with one unknown function:
\begin{enumerate}
\item
     the dilaton field is absent: $\lambda =0, \varphi \equiv \delta \varphi
     \equiv 0$;
\item there are no extra dimensions: $\mu \equiv \delta \mu \equiv 0, N=1$;
\item some combinations of (52) and (53) decouple.
\end{enumerate}

     Due to  [18], the third possibility is realized for black-hole solutions
     when $d=1$; however, this is probably not the case for arbitrary $d$.
     Here we confine the study to the first variant when the system dynamics
     is driven by extra dimensions. We shall verify that the
     arbitrariness of $d$ does not affect the stability conclusions.

     Separating variables in (53) and transforming $\mu$ and $u$ to
     obtain the normal Liouville form of the mode equation,

\begin{equation}                                                      
     \delta \mu = \e^{i\Omega t}y(x)/r^{d'/2}, \ \ x=-\int r^{d'}(u)du,
\end{equation}
     we obtain the Schr\"odinger-like equation
\begin{equation}                                                      
     y_{xx}+[\Omega^2-V(x)]y=0
\end{equation}
     with the effective potential
\begin{equation}                                                      
      V(x) =
      \frac{2}{r^{2d'}}\left[
       \frac{(N-d)(N+1)}{d'^2}\left(\frac{r\mu'^2}{r'}\right)'
       +\frac{d'r^{d'/2}}{4}\left(\frac{r'}{r^{(d+3)/2}}\right)'
       -\frac{N-d}{d'}\mu''\right]
\end{equation}
     where  $d'=d+1$, again.

     Our static system is unstable if there exist physically allowed
     solutions of (57) with $\Omega^2<0$ (negative energy levels in the
     language of the Schr\"odinger equation).

     $V(x)$ is a rather complicated function of $x$, undetermined explicitly,
     so that Eq.(57) can hardly be solved exactly. However, the main
     conclusions on the system stability can be drawn on the basis of the
     qualitative behavior of $V(x)$ and the asymptotic forms of the solutions
     to Eq.(57). The latter can be classified according to different
     behaviors of the background static solutions for different values of the
     integration constants (the notations A-D coincide with those in Section
     3).
\vskip2mm

     A. $h<0$ or (and) $u_1<0$, the case of a repelling Reissner-Nordstr\"om
     singularity. In the limit $u\rightarrow u_{\max}$ [18]
\begin{eqnarray}                                                 
     x & \sim &(u_{\max}-u)^{2+1/N},\\
     V(x) & = &-\frac{(N+1)(3N+1)}{4(2N+1)^2 x^2}(1+o(1)).          
\end{eqnarray}
     so that $V$ has a negative pole. The boundary conditions (54), (55) take
     the form
\begin{eqnarray}
     y\rightarrow 0 \ \ \ & {\rm at}\ \ x \rightarrow \infty,\\     
     x^{-(N+1)/(4N+2)}|y|< \infty & {\rm at}\ \ x\rightarrow 0.      
\end{eqnarray}
     At spatial infinity ($x\rightarrow \infty$) $V(x)$ behaves like
     const$/x^{d+2}$ (in this and all the other cases).

     We see that both the potential and the boundary condition near
     $x=0$ are independent of $d$ and coincide with those derived in Refs.
     [6,18]. Therefore the conclusion is also the same. It looks as follows.

     The asymptotic forms of the solutions to Eq.(57) for $\Omega^2<0$ at
     $x\rightarrow 0$ and $x\rightarrow \infty$ are
\begin{eqnarray}
     x\rightarrow\infty:\ &y= C_1 \e^{-|\Omega|x} +C_2 \e^{|\Omega|x},\\ 
      x\rightarrow 0: &y= C_3 x^{s_1} +C_4 x^{s_2},\ \
            s_{1,2}=\frac{1}{2}\left(1 \pm \frac{N}{2N+1}\right).    
\end{eqnarray}
     The condition (63) is satisfied for any $C_3$
     and $C_4$. Therefore the solution $y(x)$ chosen on the basis of (61),
     i.e., that with $C_2=0$, is admissible and realizes the instability of
     our static system.

     In other words, in the potential well corresponding to the bare
     singularity there exist arbitrarily low ``energy levels'' for the
     perturbations.  Thus the system is unstable and the instability is of a
     catastrophic nature since the increment $|\Omega|$ has no upper bound.
\vskip2mm

     B. The case $h>0, \ u_1 >0;  \; r\rightarrow 0$ at
     $u\rightarrow\infty$, a scalar-type central singularity where the
     ``scalar field'' $\mu$ is infinite. At $u\rightarrow\infty$
\begin{equation}
     x\rightarrow 0, \ \ V(x)= -\frac{1}{4x^2}(1+o(1)).               
\end{equation}
     The boundary condition at $x\rightarrow 0$ is
\begin{equation}
     |y|/(\sqrt{x}\ln x)<\infty                                       
\end{equation}
     while the solution has the following asymptotic form at $x\rightarrow
     0$:
\begin{equation}
     y= \sqrt{x}(C_5 + C_6\ln x).                                     
\end{equation}
     Again both the potential and the boundary conditions are
     $d$-independent. Considerations similar to those in item A lead to the
     same conclusion, namely, that the system is catastrophically unstable.
\vskip2mm

     C. For the $(d+3)$-dimensional section of the D-dimensional metric there
     is a range of integration constant values when the geometry has the form
     of a ``space pocket'' (see item C in Section 3). On the contrary, in our
     $(d+3)$-dimensional metric (42), (43), containing an additional conformal
     factor, such a possibility is not realized. Indeed, using the
     expression (43) for $\beta = \ln r$ at the asymptotic
     $u\rightarrow\infty$ and the relation (26) among the constants, one can
     verify that $r$ can tend either to zero, or to a nonzero finite value
     corresponding to a black hole.
\vskip2mm

     D. Let us now consider perturbations of multidimensional black
     holes described by (20)-(24) under the conditions (31).
     The effective potential (59) may be written in an explicit form using
     the transformation (28) in terms of the $R$ coordinate:
$$                                                                    
     V(x) = V_1(x) + V_2(x), $$
$$   V_1(x)=\frac{d+1}{2dr^{2(d+1)}}\frac{R-2k}{(R+p)^2}
          \{(R+pN_-)[R(2k+p-pN_-)+2kpN_-]\; \; \; \;   $$
$$    \; \; \; \; +(R-2k)[Rp(1-N_-)+\frac{d-1}{2d}(R+pN_-)^2]\}, $$
\begin{equation}
     V_2(x)=\frac{2pN_- (2k+pN_-)R(R-2k)}{(R+pN_-)^2 r^{2(d+1)}}.
\end{equation}
     where $N_- = (N-d)/[N(d+1)]$.

     As follows from (68), $V_1>0$ and $V_2 > 0$ when $R>2k$.
     The boundary conditions (54), (55) correspond to those conventional in
     quantum mechanics and thus the positiveness of $V(x)$ means that
     solutions to (57) with $\Omega^2<0$ are absent.
     Consequently, multidimensional BHs are stable under monopole
     perturbations. Other types of multidimensional spherically symmetric
     solutions are strongly unstable.

     This generalizes the conclusions of Refs. [6,18] for the case of
     arbitrary $d$.
\section{CONCLUSIONS}

     We have seen that multidimensional black holes are stable
     under monopole perturbations, while other types of (electro)vacuum
     spherically symmetric solutions are strongly unstable.  This property
     distinguishes the black hole (BH) solutions from all
     the other possible solutions
     to the multidimensional field equations and supports the view that
     even if space, time and gravity are described by some
     multidimensional model, realistic collapse of isolated bodies should
     lead to black hole formation, as it is conventionally asserted in
     general relativity. In particular, this favours the models with a
     great number of primordial BHs present in the Early Universe, at
     epochs when there was no crucial distinction between the physical and
     internal dimensions.

\bigskip

\bigskip

\noindent
{\bf ACKNOWLEDGMENT}\\
The work was sponsored by KAI e.V. Berlin through the WIP project
016659/p and partly by the Russian Ministry of Science. One of us (V. M.)
was supported by DFG grant 436 RUS 113-7-2.
V. M. also thanks the colleagues of WIP gravitation project group at Potsdam
University for their hospitality.


\pagebreak
\noindent
{\bf REFERENCES}

\begin{description}
\item[1.]  T.Kaluza, {\it Preuss.Akad.Wiss.} (1921), 966.
\item[2.]  O.Klein, {\it Z.Phys.} {\bf 37} (1926), 895.
\item[3.]  M.Green, I.Schwarz and E.Witten, ``Superstring Theory'',
          Cambridge Univ.Press, 1986.
\item[4.]  V.N.Melnikov, {\it in}
          ``Results of Science and Technology. Gravitation and Cosmology''
          (V.N.Melnikov, Ed.), Vol.1, p.49, VINITI Publ.,
          Moscow 1991 (in Russian). Preprint CBPF-NF-051/93, Rio de
          Janeiro, Brasil.
\item[5.]  K.A.Bronnikov, V.D.Ivashchuk and
          V.N.Melnikov, {\it Nuovo Cim.} {\bf 102} (1988), 209.
\item[6.]  K.A.Bronnikov and V.N.Melnikov, {\it in}
          ``Results of Science and Technology. Gravitation and Cosmology''
          (V.N.Melnikov, Ed.), Vol.4, p.67,
          VINITI Publ., Moscow 1992 (in Russian).
          V.N. Melnikov, Preprint CBPF-NF-051/93, Rio de Janeiro,
           Brasil.
\item[7.]  K.A.Bronnikov, {\it Ann.der Phys.(Leipzig)} {\bf 48} (1990),
          527.\\
           S.B. Fadeev, V.D. Ivashchuk, and V.N. Melnikov, {\it Phys.
           Lett.} {\bf A161}(1991) 98.
\item[8.]  S.B.Fadeev, V.D.Ivashchuk and V.N.Melnikov, {\it Chinese
          Phys.Lett.} {\bf 8} (1991), 439. In: {\it Gravitation and Modern
            Cosmology}, Plenum Publ., NY (1991) 37.
\item[9.]  F.R.Tangherlini, {\it Nuovo Cim.} {\bf 27} (1963), 636.
\item[10.]   R.Myers, {\it Phys.Rev.D} {\bf 34} (1986), 1021.
\item[11.]   K.A.Bronnikov, {\it Acta Phys.Polon.} {\bf B4} (1973),
             251.\\
             K.P. Stanukovich, and V.N. Melnikov, {\it Hydrodynamics,
             Fields, and Constants in the Theory of Gravitation}. Moscow,
             Energoatomizdat, 1983 (in Russian).

\item[12.]   K.A.Bronnikov and G.N.Shikin, {\it Izvestiya Vuzov, Fizika}
          (1977), No.9, 25 (in Russian).
\item[13.]   K.A.Bronnikov, {\it Izvestiya Vuzov, Fizika} (1991), No.7,
          24 (in Russian).
\item[14.]   K.A.Bronnikov, V.N.Melnikov, G.N.Shikin and
          K.P.Stanuikovich, {\it Ann.Phys. (N.Y.)} {\bf 118} (1979), 84.
\item[15.]   K.Shiraishi, {\it Mod. Phys. Lett. A} {\bf 7} (1992), 3569.
\item[16.]   O.Heinrich, {\it Astron. Nachr.} {\bf 309} (1988), 249.
\item[17.]   K.A.Bronnikov and A.V.Khodunov, {\it Gen.Rel.and Grav.} {\bf
          11} (1979), 13.
\item[18.]   K.A.Bronnikov, {\it Izvestiya Vuzov, Fizika} (1992), No.1,
          106.

\end{description}

\end{document}